# EFFECTS OF DISORDER WITH FINITE RANGE ON THE PROPERTIES OF $D$-WAVE SUPERCONDUCTORS

Carsten T. Rieck, Kurt Scharnberg, and Simon Scheffler
*I. Institut für Theoretische Physik,*
*Fachbereich Physik der Universität Hamburg,*
scharnberg@physnet.uni-hamburg.de

**Abstract**    It has long been established that disorder has profound effects on unconventional superconductors and it has been suggested repeatedly that observation and analysis of these disorder effects can help to identify the order parameter symmetry. In much of the relevant literature, including very sophisticated calculations of interference and weak localization effects, the disorder is represented by $\delta$-function scatterers of arbitrary strength. One obvious shortcoming of this approximation is that resonant scattering resulting from the wavelength of the scattered quasiparticle matching the spatial extent of the defect is not included. We find that the mitigation of the Tc-reduction, expected when d-wave scattering is included, is very sensitive to the average strength of the scattering potential and is most effective for weak scatterers. Disorder with finite range not only has drastic effects on the predicted density of states at low energies, relevant for transport properties, but affects the spectral function at all energies up to the order parameter amplitude. The gap structure, which does not appear to be of the simplest d-wave form, should show a defect-dependent variation with temperature, which could be detected in ARPES experiments.

## Introduction

The driving force behind the study of disorder effects in superconductors is the hope that such investigations will give information on the pairing state and the pairing interaction. Indeed, qualitative differences are expected between conventional and unconventional superconductors, the latter being defined by a vanishing Fermi surface average of the order parameter. Potential scattering in an anisotropic conventional superconductor would at high enough concentration lead to a finite, isotropic gap, while unconventional superconductors would acquire midgap states before superconductivity is destroyed. Conven-tional superconductors show this kind of behavior in the presence of spin-flip scattering. [Abrikosov and Gor'kov, 1961] Because of the innate magnetism in high temperature superconductors, nonmagnetic impurities can induce local moments and thus blur the seemingly clear distinction between potential and spin-flip scattering.

Here, we shall assume that we are dealing with $d$-wave superconductors. Since for unconventional superconductors there is no qualitative difference between these two types of scattering, we shall confine ourselves to the study of potential scattering. Even with this limitation there is a wide range of theoretical predictions as regards $T_c$-suppression, density of states, transport properties etc, depending on the way disorder is modelled and depending on the analytical and numerical approximations employed to derive experimentally verifiable conclusions. [Atkinson et al., 2000]

In much of the published work, the scattering centers are assumed to be short ranged ($s$-wave scattering). Even for defects within the $CuO_2$-planes it seems rather doubtful, that their effect is limited to a single site. Defects due to oxygen nonstoichiometry and cation disorder, which reside on lattice sites away from the conducting $CuO_2$-planes, are only poorly screened and hence are certainly long ranged. We generalize the selfconsistent T-matrix approximation (SCTMA) to scattering potentials of arbitrary range and then calculate single particle selfenergy correction in 2D $d$-wave superconductors. First application is to the $T_c$-suppression and the angle dependent density of states as seen in ARPES.

## 1. Model assumptions

We start from a two-dimensional free electron gas. The Fermi surface in this model is circular and the band width is infinite. Superconducting properties are described in a Fermi surface restricted approach, so that the weak coupling selfconsistency equation reads

$$\Delta(\varphi) = \pi N_F T \sum_{\omega_n} \int \frac{d\psi}{2\pi} \mathcal{V}(\varphi, \psi) \, g^1(\psi, i\omega_n) \,, \tag{1}$$

where $g^1(\psi, i\omega_n)$ is the energy integrated anomalous Green function (8). If the pairing interaction is separable such that $\Delta(\varphi) = \Delta(T) e(\varphi)$, the amplitude and the ratio $2\Delta(0)/T_c$ depend sensitively on details of disorder. The variation with $\varphi$, however, is completely unaffected by the introduction of disorder. This changes when the pairing interaction is nonseparable as in the spin fluctuation model.[Monthoux and Pines, 1993; Dahm et al., 1993] To elucidate the effects of disorder it is sufficient to consider a two-component order parameter

$$\Delta(\varphi) = \Delta_2 \, 2\cos 2\varphi + \Delta_6 \, 2\cos 6\varphi \,. \tag{2}$$

The pairing interaction that gives such an order parameter is of the form

$$N_F \, \mathcal{V}(\varphi, \psi) = -2 \sum_{\{\nu,\mu\}=\{2,6\}} \cos \nu\varphi \, \Lambda_{\nu\mu} \, \cos \mu\psi \, . \tag{3}$$

This could be diagonalized to give two states, each with its own transition temperature, when both eigenvalues are positive. We shall show that momentum dependent scattering mixes the two eigenstates even at $T_c$.

For a specific example of an impurity potential with finite range we choose a Gaussian. Taking matrix elements between plain wave states $\vec{k}_F$ and $\vec{k}'_F$ gives

$$v(\varphi) = v_0 \, \frac{e^{\gamma \cos \varphi}}{I_0(\gamma)} = v_0 \sum_{n=-\infty}^{+\infty} \frac{I_n(\gamma)}{I_0(\gamma)} \cos n\varphi = v_0 \sum_{n=-\infty}^{+\infty} u_n \, e^{in\varphi} \, . \tag{4}$$

$\varphi$ is the angle between $\vec{k}_F$ and $\vec{k}'_F$. $v_0$ is the average of $v(\varphi)$ over the Fermi surface so that $u_0 = 1$. Below, we parametrize $v_0$ in terms of an $s$-wave scattering phase shift $\delta_0$: $\pi N_F v_0 = \tan \delta_0$. The last equality in (4) defines the expansion coefficients $u_n$, which are often treated as free parameters.

## 2. Selfconsistent $T$−matrix approximation (SCTMA)

Within the Fermi surface restricted approach, the impurity averaged Green function has the form

$$\hat{G}(\boldsymbol{k}, \omega) = \left[ \omega \hat{\sigma}_0 - \varepsilon(\boldsymbol{k}) \hat{\sigma}_3 - \Delta(\varphi) \hat{\sigma}_1 - \hat{\Sigma}(\varphi, \omega) \right]^{-1} \, , \tag{5}$$

where $\hat{\Sigma}(\varphi, \omega) = n_{\mathrm{imp}} \, \hat{t}(\varphi, \varphi; \omega)$ is proportional to the single defect $\hat{T}$–matrix:

$$\hat{t}(\varphi, \phi; \omega) = v(\varphi - \phi)\hat{\sigma}_3 + \pi N_F \int \frac{d\psi}{2\pi} v(\varphi - \psi) \, \hat{\sigma}_3 \, \hat{g}(\psi, \omega) \, \hat{t}(\psi, \phi; \omega) \, . \tag{6}$$

Expanding $\hat{t}$ and $\hat{g}$ in terms of Pauli matrices $\sigma^\ell$ and the unit matrix we obtain four coupled one-dimensional integral equations for the components $t^\ell(\varphi, \phi)$ of the matrix $\hat{t}$, which contain the energy integrated normal and anomalous retarded Green functions

$$g^0(\psi, \omega_+) = -\frac{\omega - \Sigma^0(\psi, \omega_+)}{\sqrt{[\Delta(\psi) + \Sigma^1(\psi, \omega_+)]^2 - [\omega - \Sigma^0(\psi, \omega_+)]^2}} \tag{7}$$

$$g^1(\psi, \omega_+) = -\frac{\Delta(\psi) + \Sigma^1(\psi, \omega_+)}{\sqrt{[\Delta(\psi) + \Sigma^1(\psi, \omega_+)]^2 - [\omega - \Sigma^0(\psi, \omega_+)]^2}} \tag{8}$$

$g^3(\psi, \omega_+)$ is assumed to vanish, because it is an odd function of energy.
The $t^\ell$'s are expanded in Fourier series' and the coefficients are collected in the form of matrices $\tilde{t}^\ell$ with elements:

$$t_{nm}^\ell = \pi N_F \int_0^{2\pi} \frac{d\varphi}{2\pi} \int_0^{2\pi} \frac{d\phi}{2\pi} \, t^\ell(\varphi, \phi) \, e^{-in\varphi + im\phi} \, . \tag{9}$$

For an order parameter with $d_{x^2-y^2}$–symmetry, $g^\ell(\psi;\omega)$ has the general form

$$g^\ell(\psi;\omega) = \sum_{m=-\infty}^{+\infty} g^\ell_{q,q+|4m-2\ell|}(\omega) \cos\left[(4m-2\ell)\psi\right] \qquad \ell = 0, 1. \quad (10)$$

The expansion coefficients are independent of $q$, but have been written in this form to define matrices $\tilde{g}^\ell$ corresponding to $\tilde{t}^\ell$. When the Fourier coefficients $u_n \tan \delta_0$ of the potential (4) are written in the form of a diagonal matrix $\tilde{v}$, the four integral equations are transformed to four equations for $\tilde{t}^\ell$. From these, $\tilde{t}^2$ and $\tilde{t}^3$ can be eliminated, so that we finally obtain.

$$\tilde{t}^0 \pm \tilde{t}^1 = \left[1 - \tilde{v}\left(\tilde{g}^0 \mp \tilde{g}^1\right)\tilde{v}\left(\tilde{g}^0 \pm \tilde{g}^1\right)\right]^{-1} \tilde{v}\left(\tilde{g}^0 \mp \tilde{g}^1\right)\tilde{v} \quad (11)$$

In general, a selfconsistent solution of these two equations can only be obtained numerically.

## 3.   Transition Temperature $T_c$

In order to calculate $T_c$, the selfconsistency equation (1) is linearized, which implies that the Green functions (7,8) are to be replaced by

$$g^0 = -i \, \text{sgn}\, \omega_n \quad \text{and} \quad g^1(\psi, i\omega_n) = -\frac{\Delta(\psi) + \Sigma^1(\psi, i\omega_n)}{|i\omega_n - \Sigma^0(\psi, i\omega_n)|}. \quad (12)$$

$\Sigma^0$ is required to zeroth order in the order parameter, so that $\tilde{t}^0$ in (11) is diagonal and both $\tilde{t}^1$ and $\tilde{g}^1$ can be neglected. Together with (9) we obtain

$$\Sigma^0(i\omega_n) = -i\,\text{sgn}\,\omega_n\, \Gamma_N^{\text{el}} \sum_{m=-\infty}^{\infty} \frac{u_m^2}{\cos^2 \delta_0 + u_m^2 \sin^2 \delta_0}. \quad (13)$$

Here, $\Gamma_N^{\text{el}} = \frac{n_{\text{imp}}}{\pi N_F} \sin^2 \delta_0$ is a scattering rate that determines the impurity limited normal state d.c. conductivity.[Hensen et al., 1997] This normal state selfenergy is isotropic, as was to be expected because in our model rotational symmetry is broken only by the superconducting order. Note, that taking the unitary limit $\delta_0 \to \pi/2$ would cause problems with the convergence of this series.
$\hat{t}^1$ and $\hat{g}^1$ are non-diagonal, but since we can neglect products of these matrices we obtain equally easily (11)

$$\Sigma^1(\psi, i\omega_n) = \sum_{\ell=-\infty}^{\infty} g^1_{q,q+4\ell-2}\, \lambda^1_{4\ell-2}\, \cos\left[(4\ell-2)\psi\right], \quad (14)$$

where

$$\lambda^1_{4\ell-2} = -\Gamma_N^{\text{el}} \sum_{m=-\infty}^{\infty} \frac{u_m u_{m+4\ell-2}\left(1 + u_m u_{m+4\ell-2}\tan^2\delta_0\right)}{(\cos^2\delta_0 + u_m^2 \sin^2\delta_0)(1 + u_{m+4\ell-2}^2 \tan^2\delta_0)}. \quad (15)$$

This, too, would seem to diverge for $\delta_0 = \pi/2$. From (12) we find by inserting the expansions (1), (8), (13), (14):

$$g^1_{q,q+4\ell-2} = -\frac{\Delta_{4\ell-2}}{|\omega_n| + \pi T_c \lambda_{4\ell-2}}, \qquad (16)$$

where the pair breaking parameters are given by

$$\lambda_{4\ell-2} = \frac{\Gamma^{\text{el}}_N}{\pi T_c} \frac{1}{2} \sum_{m=-\infty}^{\infty} \frac{(u_m - u_{m+4\ell-2})^2}{(\cos^2 \delta_0 + u_m^2 \sin^2 \delta_0)(1 + u_{m+4\ell-2}^2 \tan^2 \delta_0)}. \qquad (17)$$

**Special cases:** For an isotropic order parameter, $u_{m+4\ell-2}$ has to be replaced by $u_m$ and pair breaking from defect scattering is absent, irrespective of the exact form of the scattering potential. For $\delta-$functions scatterers, only $v_0 \neq 0$. Then the pair breaking is equally effective whatever the exact form of the $d_{x^2-y^2}$ order parameter: $\lambda_{4\ell-2} = \frac{1}{\pi T_c} \Gamma^{\text{el}}_N$.

Taking the unitary limit in (17) it would appear as if every term in the series vanishes. However, when the series is terminated at $m = \pm m_0$ [Kulić and Dolgov, 1999], one finds $\lambda_{4\ell-2} = \frac{1}{\pi T_c} \Gamma^{\text{el}}_N (4\ell - 2)$, independent of $m_0$. The approach to this limit, however, is sensitive to the choice of $m_0$. In summary, strong scatterers are more effective in breaking pairs when they have a finite range.

In the Born limit, (17) can be expressed in terms of the square of the potential:

$$\lambda^{\text{Born}}_{4\ell-2} = \frac{\Gamma}{\pi T_c} (\pi N_F)^2 \int_0^{2\pi} \frac{d\varphi}{2\pi} v^2(\varphi) \{ 1 - \cos([4\ell-2]\varphi) \} > 0. \qquad (18)$$

This can become arbitrarily small and would vanish for $v^2(\varphi) \propto \delta(\varphi)$.

For the Gaussian model potential (4) the integral can be evaluated

$$\lambda^{\text{Born}}_{4\ell-2} = \frac{\Gamma^{\text{el}}_N}{\pi T_c} \frac{1}{I_0^2(\gamma)} [I_0(2\gamma) - I_{4\ell-2}(2\gamma)]. \qquad (19)$$

If we had normalized the potential such that the average of $v^2(\varphi)$ were kept constant, this would be a monotonically decreasing function of $\gamma$, approaching zero for $\gamma \to \infty$. When only a few terms are kept in the Born limit of Eq. (17), a qualitatively different behavior can result. This discrepancy arises because in the Born limit $v^2(\varphi)$ rather than $v(\varphi)$ is approximated.

In order to calculate $T_c$, we choose the pairing interaction (3). This leads to two coupled linear equations for the two order parameter amplitudes $\Delta_2$ and $\Delta_6$ in (2). From these we obtain $\Delta_6/\Delta_2$ and two real solutions for $T_c$:

$$\ln \frac{\omega_D}{2\pi T_c} = \frac{1}{2} \left[ \frac{1}{\lambda_+} + \frac{1}{\lambda_-} + \psi_2 + \psi_6 \right] \pm \frac{1}{2} \left[ \left( \frac{1}{\lambda_+} - \frac{1}{\lambda_-} \right)^2 + \right. \qquad (20)$$

$$\left. + (\psi_2 - \psi_6)^2 - 2(\psi_2 - \psi_6) \left( \frac{1}{\lambda_+} + \frac{1}{\lambda_-} \right) \frac{\Lambda_{22} - \Lambda_{66}}{\Lambda_{22} + \Lambda_{66}} \right]^{\frac{1}{2}}.$$

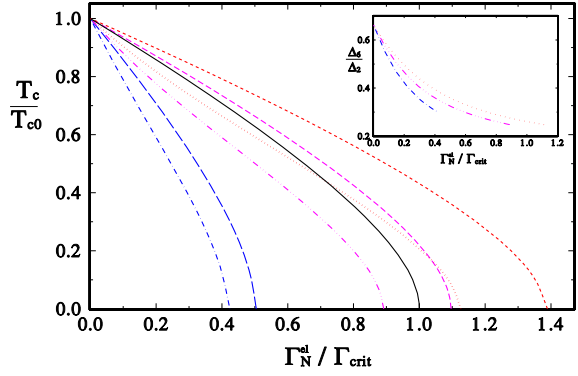

*Figure 1* Critical temperature versus $\Gamma_N^{el}$ for the Gaussian potential with width $\gamma = 5$ for $\delta_0 = 0.05\pi, 0.3\pi, 0.5\pi$. full line: Abrikosov-Gorkov, dashed lines: one-component OP, dot-(dash) lines: two-component OP (2). Inset: $\Delta_6/\Delta_2$ versus $\Gamma_N^{el}$ for the same three values of $\delta_0$ with the parameters of the pairing interaction chosen such that $T_c^+ = 90$ K, $T_c^- = 30$ K and $\Lambda_{26}/\Lambda_{22} = 0.1$.

with $\psi_i = \psi\left(\frac{1}{2} + \frac{\lambda_i}{2}\right)$. $\lambda_\pm = \frac{1}{2}(\Lambda_{22} + \Lambda_{66}) \pm \frac{1}{2}\sqrt{(\Lambda_{22} - \Lambda_{66})^2 + 4\Lambda_{26}^2}$ are the eigenvalues of the matrix $(\Lambda_{\nu\mu})$ Only when both $\lambda_+$ and $\lambda_-$ are positive, do we have two transition temperatures. The solution found for $\lambda_- < 0$ is an artefact. For the case of pure $s$−wave scattering this equation reduces to

$$\ln\frac{\omega_D}{2\pi T_c} = \psi\left(\frac{1}{2} + \frac{\Gamma_N^{el}}{2\pi T_c}\right) + \frac{1}{\lambda_\pm}. \quad (21)$$

This is the famous and well-known Abrikosov-Gorkov result. [Abrikosov and Gor'kov, 1961] We have now shown that this is independent of the exact form of the order parameter when the scattering is isotropic.

In the limit $\lambda_- \to 0^+$, (20) reduces to

$$\ln\frac{T_c}{T_{c0}} = \frac{\Lambda_{22}}{\Lambda_{22} + \Lambda_{66}}\left[\psi\left(\frac{1}{2}\right) - \psi_2\right] + \frac{\Lambda_{66}}{\Lambda_{22} + \Lambda_{66}}\left[\psi\left(\frac{1}{2}\right) - \psi_6\right]. \quad (22)$$

In both these special cases, $\Delta_6/\Delta_2$ is independent of disorder.[Haas et al., 1997] A variation of $\Delta_6/\Delta_2$ can, therefore, only be expected for $\lambda_2 \neq \lambda_6$ and $\Lambda_{26}^2 \neq \Lambda_{22}\Lambda_{66}$. However, since $\Delta_6/\Delta_2$ depends only on $\psi_2 - \psi_6$ the variation remains modest even for $T_c \to 0$. The inset of Fig. 1 gives an example of the variation of $\Delta_6/\Delta_2$ with $\Gamma_N^{el}$ and $\delta_0$. The main panel shows the transition temperature normalized to its clean limit value as function of the scattering rate, normalized to the critical scattering rate for pure $s$-wave scattering $\Gamma_{crit} = 0.882 T_{c0}$, for three values of the scattering phase shift. In the case of the single component order parameter $\propto \cos 2\varphi$, $\ln\frac{T_c}{T_{c0}} = \psi\left(\frac{1}{2}\right) - \psi\left(\frac{1}{2} + \frac{\lambda_2}{2}\right)$ (dashed lines) shows deviations from (21) that were to be expected from the variation of the pair breaking parameter $\lambda_2$ with $\delta_0$, which arises from the finite range of the scattering potential. The reduction in slope observed in electron-irradiated samples [Rullier-Albenque et al., 2003] is compatible only with the assumption of weak scattering. By the same token one concludes that Zn-impurities are strong scatterers. The regime of linear variation can be increased only if the order parameter is some general basis function of the $B_1$ irreducible representation of the point group $C_{4v}$ and if the scattering is weak. In the example shown in Fig. 1 this change is accompanied by an increase in slope. Finally, we note that an equation formally identical to (22) has been obtained in the Born approximation for arbitrary singlet order parameters and an anisotropic scattering potential of the form $v(\varphi, \phi) = v_i + v_a f(\varphi) f(\phi)$ [Haran̄ and Nagi, 1996].

## 4. Density of States

The normalized angle dependent DoS, which is measured in ARPES experiments [Borisenko et al., 2002], can be obtained directly from (7): $\frac{N(\omega, \varphi)}{N_F} = -\mathcal{I}m\, g^0(\varphi, \omega_+)$. Results for strong scatterers and a single component OP $\propto \cos 2\varphi$ are shown in Fig. 2.

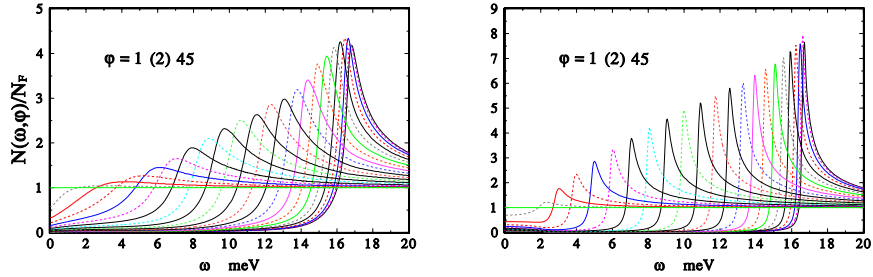

*Figure 2* Left panel: Gaussian potential with $\gamma = 5$. Right panel: Point like scatterers. Note the difference in scale. $\delta_0 = 0.49\pi$, $\Gamma_N^{el} = 0.2$ meV

In the clean limit, we would have square root singularities at $\omega = \Delta \cos 2\varphi$. Defects causing pronounced forward scattering broaden and reduce these singularities much more strongly than point like scatterers. The leading edge gap (LEG), as measured in ARPES, is much less clearly defined. The angle dependence of the LEG, differs substantially from the angle dependence $\cos 2\varphi$ of the OP, being rather $U$-shaped near the node. In this respect, forward scattering would have the same effect as the admixture of a $\cos 6\varphi$ component to the order parameter.

At very low frequencies, the angle integrated DoS is reduced by invoking forward scattering. Details depend sensitively on the choice of the coefficients $v_n$ in (4). This change in the DoS would greatly affect the thermal conductivity and the microwave conductivity.

## 5. Summary

In the kind of theory developed here, the observed variation in the slope of the disorder-induced $T_c$-degradation can be attributed to different strengths of the scattering potentials. A mitigation of this $T_c$-degradation occurs only for strong forward scattering by weak potentials. This could explain why high temperature superconductivity is rather insensitive to cation disorder and disorder associated with oxygen non-stoechiometry. A qualitative change of the dependence of $T_c$ on the scattering rate is found only when the $d$-wave order parameter has a more complicated angular dependence than $\cos[(4m - 2\ell)\varphi]$. The angle dependent DoS reflects details of the defect potential which could be measured for varying defect concentration $n_{\text{imp}}$ by ARPES. Sufficiently low temperatures to eliminate inelastic scattering and high resolution are, of course a precondition. Comparison with $T_c(n_{\text{imp}})$ would provide clues to the form of the pairing interaction $\mathcal{V}(\varphi, \psi)$.